# High Performance Computing for Geospatial Applications: A Retrospective View

Marc P. Armstrong
The University of Iowa
marc-armstrong@uiowa.edu

## Abstract

Many types of geospatial analyses are computationally complex, involving, for example, solution processes that require numerous iterations or combinatorial comparisons. This complexity has motivated the application of high performance computing (HPC) to a variety of geospatial problems. In many instances, HPC assumes even greater importance because complexity interacts with rapidly growing volumes of geospatial information to further impede analysis and display. This chapter briefly reviews the underlying need for HPC in geospatial applications and describes different approaches to past implementations. Many of these applications were developed using hardware systems that had a relatively short life-span and were implemented in software that was not easily portable. More promising recent approaches have turned to the use of distributed resources that includes cyberinfrastructure as well as cloud and fog computing.

## 1.0 Introduction

High performance computing (HPC) has been used to address geospatial problems for several decades (see, for example, Sandu and Marble 1988; Franklin *et al*. 1989; Mower 1992). An original motivation for seeking performance improvements was the intrinsic computational complexity of geospatial analyses, particularly combinatorial optimization problems (Armstrong 2000). Non-trivial examples of such problems require considerable amounts of memory and processing time, and even now, remain intractable. Other spatial analysis methods also require substantial amounts of computation to generate solutions. This chapter briefly reviews the computational complexity of different kinds of geospatial analyses and traces the ways in which HPC has been used in the past and during the present era. Several HPC approaches have been investigated, with developments shifting from an early focus on manufacturer-specific systems, which in most cases had idiosyncrasies (such as parallel language extensions) that limited portability. This limitation was recognized and addressed by using software environments that were designed to free programmers from this type of system dependence (e.g., Message Passing Interface). This is also acknowledged for example, by the work presented in Healey *et al*. (1998) with its focus on algorithms rather than implementations. Later approaches used flexible configurations of commodity components linked using high-performance networks. In the final section of this chapter, still-emerging approaches such as cloud and edge computing are briefly described.

## 2.0 Computational Complexity of Geospatial Analyses: A Brief Look

Computing solutions to geospatial problems, even relatively small ones, often requires substantial amounts of processing time. In its most basic form, time complexity considers how the number of executed instructions interacts with the number of data elements in an analysis. Best case, average and worst case scenarios are sometimes described, with the worst case normally reported. In describing complexity, it is normative to use "Big $O$" notation, where $O$ represents the order of, as in the algorithm executes on the order of $n^2$, or $O(n^2)$. In geospatial data operations it is common to encounter algorithms that have an order of at least $O(n^2)$ because such complexity obtains for nested loops, though observed complexity is slightly less since diagonal values are often not computed (the distance from a point to itself is zero). Nevertheless, the complexity remains $O(n^2)$, or quadratic, because the $n^2$ factor is controlling the limit. In practice, any complexity that is $n^2$ or worse becomes intractable for large problems. Table 1 provides some simple examples to demonstrate the explosive growth in computational requirements of different orders of time complexity. The remaining parts of this section sketch out some additional examples of complexity for different kinds of geospatial analyses.

Table 1. Number of operations required with variable orders of time complexity and problems sizes.

| Input | $\log n$ | $n \log n$ | $n^2$ | $n^3$ | $2^n$ | $n!$ |
|-------|-------|--------|-----|------|---------|----------|
| 1 | 0 | 1 | 1 | 1 | 2 | 1 |
| 2 | 0.301 | 0.602 | 4 | 8 | 4 | 2 |
| 3 | 0.477 | 1.431 | 9 | 27 | 8 | 6 |
| 4 | 0.602 | 2.408 | 16 | 64 | 16 | 24 |
| 5 | 0.699 | 3.495 | 25 | 125 | 32 | 120 |
| 6 | 0.778 | 4.668 | 36 | 216 | 64 | 720 |
| 7 | 0.845 | 5.915 | 49 | 343 | 128 | 5,040 |
| 8 | 0.903 | 7.224 | 64 | 512 | 256 | 40,320 |
| 9 | 0.954 | 8.586 | 81 | 729 | 512 | 36,2880 |
| 10 | 1 | 10 | 100 | 1,000 | 1,024 | 3,628,800 |
| 20 | 1.301 | 26.021 | 400 | 8,000 | 1,048,576 | 2.40E+17 |

- Spatial optimization problems impose extreme computational burdens. Consider the $p$-median problem in which $p$ facility locations are selected from $n$ candidate demand sites such that distances between facilities and demand location are minimized. In its most basic form, a brute force search for a solution requires the evaluation of $n! / [(n-p)! p!]$ alternatives (see Table 2 for some examples).
- As a result of this explosive combinatorial complexity, much effort has been expended to develop robust heuristics that reduce the computational complexity of search spaces. Densham and Armstrong (1994) describe the use of the Teitz and Bart vertex substitution heuristic for two case studies in India. This algorithm has a worst case complexity of $O(p*n^2)$. In one of their problems 130 facilities are located at 1,664 candidate sites and in a larger problem, 2,500 facilities are located among 30,000 candidates. The smaller problem required the evaluation of 199,420 substitutions per iteration, while the larger problem required 68,750,000 evaluations. Thus, a problem that was approximately 19 times larger required 344 times the number of substitutions to be evaluated during each iteration. In both cases, however, the heuristic search space was far smaller than the full universe of alternatives.

- Many geospatial methods are based on the concept of neighborhood and require the computation of distances among geographical features such as point samples and polygon centroids. For example, the $G_i^*(d)$ method (Ord and Getis 1995) requires pairwise distance computations to derive distance-based weights used in the computation of results. Armstrong, Cowles and Wang (2005) report a worst-case time complexity of $\boldsymbol{O}(n^3)$ in their implementation.

- Bayesian statistical methods that employ Markov Chain Monte Carlo (MCMC) are computationally demanding in terms of memory and compute time. MCMC samples are often derived using a Gibbs sampler or Metropolis-Hastings approach that may yield autocorrelated samples, which, in turn require larger sample sizes to make inferences. Yan et al. (2007) report, for example, that a Bayesian geostatistical model requires $\boldsymbol{O}(n^2)$ memory for $n$ observations and the Cholesky decomposition of this matrix requires $\boldsymbol{O}(n^3)$ in terms of computational requirements.

Table 2. Brute force solution size for four representative $p$-median problems.

| $n$ candidates | $p$ facilities | Possible solutions |
| --- | --- | --- |
| 10 | 3 | 120 |
| 20 | 5 | 15,504 |
| 50 | 10 | 10,272,278,170 |
| 100 | 15 | 253,338,471,349,988,640 |

These examples are only the proverbial tip-of-the-iceberg. Computational burdens are exacerbated when many methods (*e.g.*, interpolation) are used with big data, a condition that is now routinely encountered. Big data collections also introduce substantial latency penalties during input-output operations.

**3.0 Performance Evaluation**

Until recently, computational performance routinely and steadily increased as a consequence of Moore's Law and Dennard Scaling. These improvements also applied to the individual components of parallel systems. But those technological artifacts are the result of engineering innovations and do not reflect conceptual advances due to creation of algorithms that exploit the characteristics of problems. To address this assessment issue, performance is often measured using "speedup" which is simply a fractional measure of improvement ($t_1/t_2$) where times are normally execution times achieved using either different processors or numbers of processors. In the latter case,

*Speedup* = $t_1/t_n$

   where

   $t_1$ is sequential (one processor) time and

   $t_n$ is time required with $n$ processors.

Speedup is sometimes standardized by the number of processors used ($n$) and reported as efficiency where

*Efficiency$_n$ = Speedup$_n$/n.*

Perfect efficiencies are rarely observed, however, since normally there are parts of a program that must remain sequential.  Amdahl's Law is the theoretical maximum improvement that can be obtained using parallelism (Amdahl, 1967):

*Theoretical_Speedup* = 1/((1-*p*) + *p*/*n*),

> where
>
> *p* = proportion of the program that can be made parallel (1-*p* is the serial part) and,
>
> *n* is the number of processors.

As *n* grows, the right hand term diminishes and speedups tend to 1/(1-*p*). The consequence of Amdahl's law is that the weakest link in the code will determine the maximum parallel effectiveness. The effect of Amdahl's Law can be observed in an example reported by Rokos and Armstrong (1996), in which serial input-output comprised a large fraction of total compute time, which had a deleterious effect on overall speedup.

It should be noted that Amdahl had an axe to grind as he was in involved in the design of large mainframe uni-processor computer systems for IBM (System/360) and later Amdahl Computer, a lower-cost, plug-compatible, competitor of IBM mainframes.  In fact, Gustafson (1988: 533) reinterprets Amdahl's "law" and suggests that the computing community should overcome the "mental block" against massive parallelism imposed by a misuse of Amdahl's equation, asserting that speedup should be measured by scaling the problem to the number of processors.

**4.0 Alternative Approaches to Implementing HPC Geospatial Applications**

The earliest work on the application of HPC to geospatial analyses tended to focus first on the use of uni-processor architectures with relatively limited parallelism. Subsequent work exploited pipelining and an increased number of processors executing in parallel. Both cases, however, attempted to employ *vertical scaling* that relies on continued increases in the performance of system components, such as processor and network speeds.  These vertically scaled parallel systems (section 4.1) were expensive and thus scarce.  Moreover, companies that produced such systems usually did not stay in business very long (see Kendall Square Research, Encore Computer, Alliant Computer Systems).  A different way to think about gaining performance improvements is *horizontal scaling* in which distributed resources are integrated into a configurable system that is linked using middleware (NIST 2015).  This latter approach (section 4.2) has been championed using a variety of concepts and terms such as grid computing and cyberinfrastructure.

4.1 The Early Era: Vertical Scaling, Uni-processors, and Early Architectural Innovations

The von Neumann architecture serves as a straightforward basis for modern computer architectures that have grown increasingly complex during the past several decades. In its most simple form, a computer system accepts input from a device, operates on that input and moves the results of the operation to an output device. Inputs in the form of data and instructions are stored in memory and operations are typically performed by a central processing unit that can take many forms. Early computers used a single bus to access both data and instructions and contention for this bus slowed program execution in what is known as the von Neumann "bottleneck".  Though performance could be improved by increasing the clock speed of the processor, computer architects have also invented many other paths (*e.g.*, the so-called Harvard architecture), that exploit multiple busses, a well-developed memory hierarchy (Figure 1) and multiple cores and processors, in an attempt to overcome processing roadblocks.  In practice, most geospatial processing takes place using the limited number of cores that are available on the commodity chips used by desktop systems.  Some current generation CPUs, for example, feature six cores and 12 threads, thus enabling limited parallelism that may be exploited automatically by compilers and thus remain invisible to the average user.

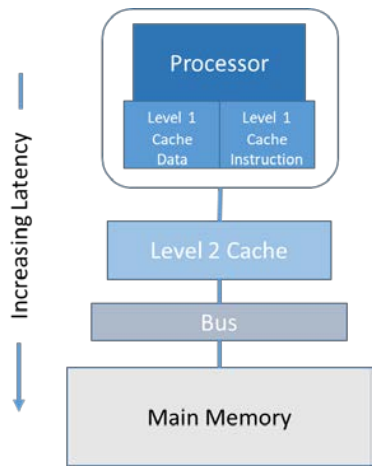

Figure 1. Memory hierarchy in an abstract representation of a uniprocessor. Level 1 cache is usually accessible in very few clock cycles, while access to other levels in the hierarchy typically requires an increased number of cycles.

When examining alternative architectures, it is instructive to use a generic taxonomy to place them into categories in a 2x2 grid with the axes representing data streams and instruction streams (Flynn 1972). The top left category (single instruction and single data streams; SISD) represents a simple von Neumann architecture. The MISD category (multiple instruction and single data streams; top right) has no modern commercial examples that can be used for illustration. The remaining two categories (the bottom row of Figure 2) are parallel architectures that have had many commercial implementations and a large number of architectural offshoots.

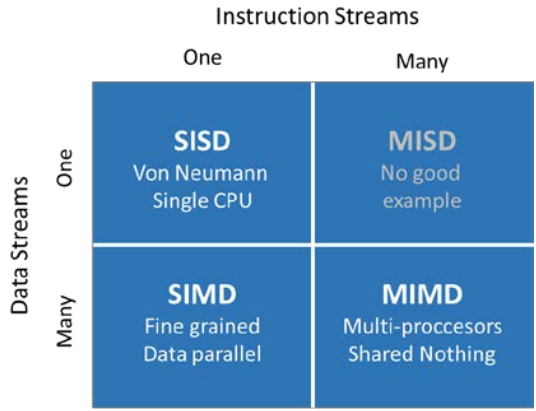

Figure 2. Flynn's taxonomy represented in a 2x2 cross-classification (S: single; M: multiple; I: instruction streams; D: data streams).

The simple SISD model (von Neumann architecture) is sometimes referred to as a scalar architecture. This is illustrated in Figure 3a, which shows that a single result is produced only after several low-level instructions have been executed.

The vector approach, as the name implies, operates on entire vectors of data. It requires the same number of steps to fill a "pipeline" but then produces a result with every clock cycle (Figure 3b). An imperfect analogy would be that a scalar approach would allow only one person on an escalator at a time, whereas a vector approach would allow a person on each step, with the net effect of much higher throughput.

Time

|  | 0 | 10 | 20 | 30 | 40 | 50 | 60 | 70 | 80 | 90 | 100 |
|---|---|---|---|---|---|---|---|---|---|---|---|
| Input | A1,B1 |  |  |  |  | A2,B2 |  |  |  |  | A3,B3 |
| Modify | A1,B1 |  |  |  |  | A2,B2 |  |  |  |  | A3,B3 |
| Add |  | A1,B1 |  |  |  |  | A2,B2 |  |  |  |  |
| Normalize |  |  | A1,B1 |  |  |  |  | A2,B2 |  |  |  |
| Round |  |  |  | A1,B1 |  |  |  |  | A2,B2 |  |  |
| Output |  |  |  |  | **A1+B1** |  |  |  |  | **A2+B2** |  |

|  | 0 | 10 | 20 | 30 | 40 | 50 | 60 | 70 | 80 | 90 | 100 |
|---|---|---|---|---|---|---|---|---|---|---|---|
| Input | A1,B1 | A2,B2 | A3,B3 | A4,B4 | A5,B5 | A6,B6 | A7,B7 | A8,B8 | A9,B9 | A10,B10 | A11,B11 |
| Modify | A1,B1 | A2,B2 | A3,B3 | A4,B4 | A5,B5 | A6,B6 | A7,B7 | A8,B8 | A9,B9 | A10,B10 | A11,B11 |
| Add |  | A1,B1 | A2,B2 | A3,B3 | A4,B4 | A5,B5 | A6,B6 | A7,B7 | A8,B8 | A9,B9 | A10, B10 |
| Normalize |  |  | A1,B1 | A2,B2 | A3,B3 | A4,B4 | A5,B5 | A6,B6 | A7,B7 | A8,B8 | A9,B9 |
| Round |  |  |  | A1,B1 | A2,B2 | A3,B3 | A4,B4 | A5,B5 | A6,B6 | A7,B7 | A8,B8 |
| Output |  |  |  |  | **A1+B1** | **A2+B2** | **A3+B3** | **A4+B4** | **A5+B5** | **A6+B6** | **A7+B7** |

Figure 3 a (top) and b (bottom). The top table shows a scalar approach that produces a single result after multiple clock cycles. The bottom table (b) shows that the same number of cycles (40) is required to compute the first result and then after that, the full pipeline produces a result at the end of every cycle. (Adapted from Karplus 1989)

Vector processing was a key performance enhancement of many early supercomputers, particularly those manufactured by Cray Research (*e.g.*, Cray X-MP) which had only limited shared memory parallelism (August *et al*. 1989). The Cray-2 was a successor to the X-MP and Griffith (1990) describes the advantages that accrue to geospatial applications with vectorized code. In particular, in the Fortran examples provided, the best improvements occur when short loops are processed; when nested do-loops are used the inner loops are efficiently vectorized, while outer loops are scalar.

In a more conventional SIMD approach to parallel computing, systems are often organized in an array (such as 64x64 = 4K) of relatively simple processors that are 4- or 8-connected.  SIMD processing is extremely efficient for gridded data because rather than cycling through a matrix element by element, all elements in an array (or a large portion of it) are processed in lockstep; a single operation (*e.g.*, add two integers) is executed on all matrix elements simultaneously.  This is often referred to as data parallelism and is particularly propitious for problems that are represented by regular geometrical tessellations, as encountered, for example, during cartographic modeling of raster data.

In other cases, while significant improvements can be observed, processing efficiency may drop because of the intrinsic sequential nature of required computation steps.  For example, Armstrong and Marciano (1995) reported substantial improvements over a then state-of-the-art workstation for a spatial statistics application using a SIMD MasPar system with 16K processors, though efficiency values were more modest.  In the current era, SIMD array-like processing is now performed using GPU (graphics processing units) accelerators (*e.g.*, Lim and Ma 2013; Tang and Feng 2017).  In short, a graphics card plays the same role that an entire computer system played in the 1990s. Tang (2017: 3196) provides an overview of the use of GPU computing for geospatial problems and correctly points out that data structures and algorithms must be transformed to accommodate SIMD architectures and that if not properly performed, only modest improvements will obtain (see also Armstrong and Marciano 1997).

MIMD computers take several forms. In general, they consist of multiple processors connected by a local high speed bus.  One simple model is the shared-memory approach, which hides much of the complexity of parallel processing from the user. This model typically does not scale well as a consequence of bus contention: all processors need to use the same bus to access the shared memory (see Figure 4).  This approach also requires that the integrity of shared data structures be maintained, since each process has equal memory modification rights.  Armstrong, Pavlik and Marciano (1994) implemented a parallel version of **G**(*d*) (Ord and Getis 1995) using an Encore Multimax shared memory system with 14 processors.  While performance improvements were observed, the results also show decreasing scalability, measured by speedup, particularly when small problems are processed, since the initialization and synchronization processes (performed sequentially), occupy a larger proportion of total compute time for those problems (Amdahl's Law strikes again).  Shekhar et al. (1996) report better scalability in their application of data partitioning to support range query applications, but, again, their Silicon Graphics shared memory system had only 16 processors.

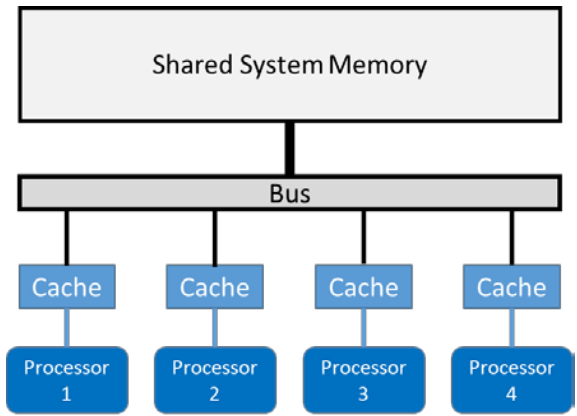

Figure 4. A simplified view of a four-processor shared memory architecture. The connection between the bus and shared memory can become a choke-point that prevents scalability.

A different approach to shared memory was taken by Kendall Square Research when implementing their KSR-1 system from the mid-1990s. Rather than employing a single monolithic shared memory it used a hierarchical set of caches (ALLCACHE, Frank *et al*. 1993). As shown in Figure 5, each processor can access its own memory as well as that of all other processing nodes to form a large virtual address space. Processors are linked by a high-speed bus divided into hierarchical zones, though the amount of time required to access different portions of shared memory from any given processor will vary if zone borders must be crossed. A processor can access its 256K sub-cache in only 2 clock cycles, it can access 32MB memory in its own local cache in 18 clock cycles, and if memory access is required on another processor within a local zone, then a penalty of 175 clock cycles is incurred (Armstrong and Marciano, 1996). However, if a memory location outside this zone must be accessed, then 600 cycles are required. This hierarchical approach exploits data locality and is a computer architectural restatement of Tobler's First Law: All things are interconnected, but near things are interconnected faster than distant things.

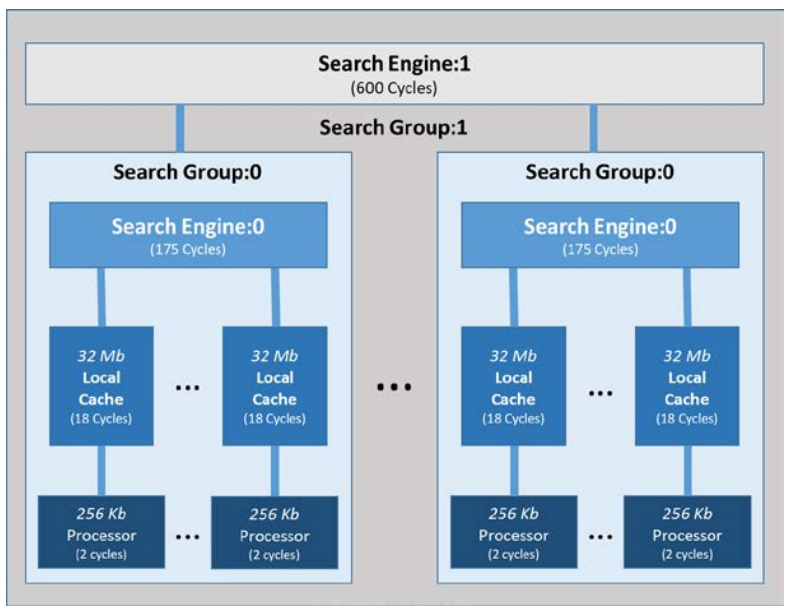

Figure 5. Non-uniform (hierarchical) memory structure of the KSR1 (after Armstrong and Marciano 1996).

## 4.2 Distributed Parallelism and Increased Horizontal Scalability

Horizontal scaling began to gain considerable traction in the early 1990s.  In contrast to the vertically-scaled systems typically provided by a single, turn-key vendor, these new approaches addressed scalability issues using a shared nothing approach that had scalability as a central design element. Stonebraker (1986), for example, described three conceptual architectures for multiprocessor systems (shared memory, shared disk and shared nothing) and concluded that shared nothing was the superior approach for database applications.

Smarr and Catlett (1992: 45) pushed this concept further in conceptualizing a metacomputer as "a network of heterogeneous, computational resources linked by software in such a way that they can be used as easily as a personal computer."  They suggest an evolutionary path in network connectivity from local area networks to wide area networks to a third stage: a transparent national network that relies on the development of standards that enable local nodes to interoperate in flexible configurations. These concepts continued to be funded by NSF and eventually evolved into what was termed grid computing (Foster and Kesselman 1999) with its central metaphor of the electric grid (with computer cycles substituting for electricity). Wang and Armstrong (2003), Armstrong, Cowles and Wang (2005) and Wang, Cowles and Armstrong (2008) illustrate the effectiveness of the grid computing paradigm to geospatial information analysis.

At around the same time, several other related concepts were being developed that bear some similarity to the grid approach.  The Network of Workstations (NOW) project originated in the mid-1990s at UC-Berkeley in an attempt to construct configurable collections of commodity workstations that are connected using what were then high-performance networks (Anderson *et al*. 1995). Armstrong and Marciano (1998) developed a NOW implementation (using Message Passing Interface, Snir *et al*. 1996) to examine its feasibility in geospatial processing (inverse-distance weighted interpolation).  While substantial reductions in computing time were realized, the processor configuration achieved only moderate levels of efficiency when more than 20 processors were used due, in part, to communication latency penalties from the master-worker approach used to assign parallel tasks.  At around the same time, Beowulf clusters were also developed with a similar architectural philosophy: commodity class processors, linked by Ethernet to construct a distributed parallel architecture.

## 4.3 Cyberinfrastructure and CyberGIS

Cyberinfrastructure is a related term that was promoted by the National Science Foundation beginning in the early 2000s (Atkins *et al*. 2003) and it continues into the present era with the establishment of NSF's Office of Advanced Cyberinfrastructure, which is part of the Computer and Information Science and Engineering Directorate. While numerous papers have described various aspects of Cyberinfrastructure, Stewart *et al*. (2010: 37) define the concept in the following way:

> "Cyberinfrastructure consists of computing systems, data storage systems, advanced instruments and data repositories, visualization environments, and people, all linked together by software and high performance networks to improve research productivity and enable breakthroughs not otherwise possible."

The "all linked together" part of the definition moves the concept from a more localized view promulgated by NOW and Beowulf to a much more decentralized, even international, scope of operation more aligned with the concepts advanced as grid computing. This linking is performed through the use of middleware, software that acts as a kind of digital plumbing to enable disparate components to work together.

The most fully realized geospatial implementation of the cyberinfrastructure concept is the CyberGIS project at the University of Illinois (Wang 2010; 2013; also see [http://cybergis.illinois.edu/)](http://cybergis.illinois.edu/)) Wang (2019) characterizes CyberGIS as a "fundamentally new GIS modality based on holistic integration of high-performance and distributed computing, data-driven knowledge discovery, visualization and visual analytics, and collaborative problem-solving and decision-making capabilities". CyberGIS uses general-purpose middleware, but goes beyond that to also implement geospatially-tailored middleware that is designed to capture spatial characteristics of problems in order to promote application specific efficiencies in locating and using distributed resources (Wang 2010). It seems clear, based on current trends, that the concept of cyberinfrastructure will continue to be central to developments in HPC at least for the next decade (NSF 2018) and that CyberGIS will continue to undergo "parallel" developments.

## 4.4 Cloud computing

Cloud computing is yet another general distributed model in which configurable computer services, such as compute cycles and storage, is provided over a network (Sugumaran and Armstrong 2017). It is, as such, a logical outgrowth of grid computing and it is sometimes referred to as utility computing.

The US National Institute of Standards and Technology has provided an authoritative definition of cloud computing (Mell and Grance 2011: 2) with five essential characteristics that are paraphrased here:

1. *On-demand self-service.* Consumers must be able to access computing capabilities, such as compute time and network storage automatically.
2. *Broad network access.* Capabilities must be available over the network and accessed through standard protocols that enable use by diverse platforms (e.g., tablets and desktop systems).
3. *Resource pooling.* Service provider's resources are pooled to serve multiple consumers, with different physical and virtual resources dynamically assigned according to consumer demand. The consumer has no control or knowledge about the location of the provided resources.
4. *Rapid elasticity.* Capabilities are elastically provided commensurate, with demand.
5. *Measured service.* Cloud systems control and optimize resource use by metering resources; usage is monitored, controlled, and reported, providing transparency to providers and consumers.

The flexibility of the cloud approach means that users and organizations are able to adapt to changes in demand. The approach also changes the economics of computing from a model that may require considerable capital investment in hardware, with associated support and upgrade (fixed) costs to one in which operational expenses are shifted to inexpensive networked clients (variable costs). While there are private clouds, cloud computing services are often public and provided by large corporate enterprises (*e.g.*, Amazon and Google) that offer attractive, tailorable hardware and software configurations. Cloud computing is reaching a mature stage and because of its tailorable cost structures and configurability, the approach will continue to be broadly adopted in the foreseeable future. Cloud computing has been demonstrated to be effective in several geospatial problem domains (Hegeman *et al*. 2014; Yang *et al*. 2013)

## 4.5 Moving closer to the edge

Despite the substantial advantages provided by cloud computing, it does suffer from some limitations, particularly latency.  Communication is, after all, limited by the speed of light and in practice it is far slower than that limit (Satyanarayanan 2017).  With the proliferation of electronic devices connected as part of the Internet of Things (estimated to be approximately 50,000,000,000 by 2020) that are generating zettabytes ($10^{21}$ bytes) of data each year, bandwidth is now a major concern (Shi and Dustdar 2016).  Trends in distributed data collection and processing are likely to persist, with one report[1] by the FCC 5G IoT Working Group suggesting that the amount of data created and processed outside a centralized center or cloud is now around 10% and will likely increase to 75% by 2022.

Communication latency is particularly problematic for real-time systems, such as augmented reality and autonomous vehicle control.  As a consequence, edge and fog computing  have emerged as important concepts in which processing is decentralized, taking place between individual devices and the cloud, to obviate the need to move massive amounts of data, and thereby increase overall computational performance.  It turns out that geography matters.  Achieving a proper balance between centralized and distributed processing is a key here.  The movement of massive amounts of data has also become a source of concern for those companies that are now providing 5G wireless network service that could be overwhelmed by data fluxes before the systems are even fully implemented.

In a fashion similar to that of cloud computing, NIST has promulgated a definition of fog computing that contains these six fundamental elements as reported by Iorga *et al*. (2018: 3-4):

1. *Contextual awareness and low latency*. Fog computing is low latency because nodes are often co-located with end-devices, and analysis and response to data generated by these devices is faster than from a centralized data center.
2. *Geographical distribution*. In contrast to centralized cloud resources, fog computing services and applications are geographically distributed.
3. *Heterogeneity*. Fog computing supports the collection and processing of different types of data acquired by multiple devices and network communication capabilities.
4. *Interoperability and federation*. Components must interoperate, and services are federated across domains.
5. *Real-time interactions*. Fog computing applications operate in real-time rather than in batch mode.
6. *Scalability and agility*. Fog computing is adaptive and supports, for example, elastic computation, resource pooling, data-load changes, and network condition variations.

The battle between centralization and de-centralization of computing is ongoing.  Much like the episodic sagas of the mainframe vs the personal computer, the cloud vs fog approach requires that trade-offs be made in order to satisfy performance objectives and meet economic constraints.  While cloud computing provides access to flexibly specified, metered, centralized resources, fog computing offloads burdens from the cloud to provide low-latency services to applications that require it.  Cloud and fog computing, therefore, should be viewed as complementary.

---

[1] https://www.fcc.gov/bureaus/oet/tac/tacdocs/reports/2018/5G-Edge-Computing-Whitepaper-v6-Final.pdf

**5.0 Summary and Conclusion**

The use of HPC in geospatial applications has had a checkered history with many implementations showing success only to see particular architectures or systems become obsolete. Nevertheless, there were lessons learned that could be passed across generations of systems and researchers.  For example, early conceptual work on geospatial domain decomposition (Armstrong and Densham 1992) informed further empirical investigations of performance that modelled processing time as a function of location (Cramer and Armstrong 1999).  A decade later, this work was significantly extended to undergird the assignment of tasks in modern cyberinfrastructure-based approaches to distributed parallelism (Wang and Armstrong 2009).

It also seems clear that HPC is reaching a maturation point with much attention now focused on the use of distributed resources that interoperate over a network (*e.g.*, cyberinfrastructure, cloud and fog computing).  Though much work remains, there is at least a clear developmental path forward (Wang, 2013).  There is a cloud on the horizon, however. Moore's Law (Moore 1965) is no longer in force (Hennessy and Patterson 2019).  As a consequence, computer architects are searching for alternative methods, such as heterogeneous processing and quantum computing, to increase performance.  It is likely, however, that many of these emerging technologies will continue to be accessed using cyberinfrastructure.

## References

Amdahl GM (1967) Validity of the single-processor approach to achieving large-scale computing capabilities. Proceedings of the American Federation of Information Processing Societies Conference (pp. 483-485). Reston, VA: AFIPS Press.

Anderson TE, Culler DE, Patterson DA, and the NOW Team (1995) A case for NOW (Networks of Workstations). IEEE Micro 15(1): 54-64

Armstrong MP (2000) Geography and computational science. Annals of the Association of American Geographers 90(1): 146–156

Armstrong MP, Densham PJ (1992) Domain decomposition for parallel processing of spatial problems. Computers, Environment and Urban Systems 16(6): 497–513

Armstrong MP, Marciano RJ (1995) Massively parallel processing of spatial statistics. International Journal of Geographical Information Systems 9(2): 169-189

Armstrong MP, Marciano RJ (1996) Local interpolation using a distributed parallel supercomputer. International Journal of Geographical Information Systems 10 (6): 713-729

Armstrong MP, Marciano RJ (1997) Massively parallel strategies for local spatial interpolation. Computers & Geosciences 23(8): 859-867

Armstrong MP, Marciano RJ (1998) A network of workstations (NOW) approach to spatial data analysis: The case of distributed parallel interpolation. Proceedings of the Eighth International Symposium on Spatial Data Handling, Burnaby, BC: International Geographical Union, pp. 287-296

Armstrong MP, Pavlik CE, Marciano RJ (1994) Parallel processing of spatial statistics. Computers & Geosciences 20(2): 91-104

Armstrong MP, Cowles M, Wang S (2005) Using a computational grid for geographic information analysis. The Professional Geographer 57(3): 365-375

Atkins DE, Droegemeier KK, Feldman SI, Garcia-Molina H, Klein ML, Messerschmitt DG, Messina P, Ostriker JP, and Wright MH (2003) Revolutionizing Science and Engineering Through Cyberinfrastructure: Report of the National Science Foundation Blue-Ribbon Advisory Panel on Cyberinfrastructure. http://www.nsf.gov/od/oci/reports/toc.jsp

August MC, Brost GM, Hsiung CC, Schiffleger AJ (1989) Cray X-MP: The birth of a supercomputer. IEEE Computer 22(1): 45-52

Cramer BE, Armstrong MP (1999) An evaluation of domain decomposition strategies for parallel spatial interpolation of surfaces. Geographical Analysis 31 (2): 148-168

Densham PJ, Armstrong MP (1994) A heterogeneous processing approach to spatial decision support systems. In Waugh TC and Healey RG (eds) Advances in GIS Research, Volume 1. London, UK: Taylor and Francis Publishers, pp. 29-45

Flynn MJ (1972) Some computer organizations and their effectiveness. IEEE Transactions on Computers C-21 (9): 948–960.  doi:10.1109/TC.1972.5009071

Foster I, Kesselman C (eds) (1999) The grid: blueprint for a new computing infrastructure. Morgan Kaufmann Publishers Inc. San Francisco, CA

Frank S, Burkhardt H, Rothnie J (1993) The KSR 1: bridging the gap between shared memory and MPPs. COMPCON Spring '93. Digest of Papers p 285-294
https://doi.ieeecomputersociety.org/10.1109/CMPCON.1993.289682

Franklin WR, Narayanaswami C, Kankanhalli M, Sun D, Zhou M-C, Wu, PYF (1989) Uniform grids: A technique for intersection detection on serial and parallel machines. In Proceedings of the Ninth International Symposium on Computer-Assisted Cartography, (Baltimore, MD 2-7 April) American Congress on Surveying and Mapping, Bethesda, MD, pp. 100-109.

Griffith DA (1990) Supercomputing and spatial statistics: a reconnaissance. The Professional Geographer 42(4): 481–92

Gustafson JL (1988) Reevaluating Amdahl's law. Communications of the Association for Computing Machinery 31 (5): 532-533

Healey R, Dowers S, Gittings B, Mineter M (eds) (1998) Parallel Processing Algorithms for GIS. Bristol, PA: Taylor & Francis.

Hegeman JW, Sardeshmukh VB, Sugumaran R, Armstrong MP (2014) Distributed LiDAR data processing in a high-memory cloud-computing environment. Annals of GIS 20(4): 255-264

Hennessy JL, Patterson DA (2019) A new golden age for computer architecture. Communications of the Association for Computing Machinery 62 (2): 48-60

Iorga M, Feldman L, Barton R, Martin MJ, Goren N, Mahmoudi C (2018) Fog Computing Conceptual Model. NIST Special Publication 500-325. Gaithersburg, MD: NIST. https://doi.org/10.6028/NIST.SP.500-325

Karplus WJ (1989) Vector processors and multiprocessors. In Hwang K, DeGroot D (eds) Parallel processing for supercomputers and artificial intelligence. New York, NY: McGraw Hill.

Lim GJ, Ma L (2013) GPU-based parallel vertex substitution algorithm for the p-median problem. Computers & Industrial Engineering 64: 381-388.

Mell P, Grance T (2011) The NIST Definition of Cloud Computing. National Institute of Standards and Technology Special Publication 800-145. Gaithersburg, MD: NIST. https://doi.org/10.6028/NIST.SP.800-145

Moore G (1965) Cramming more components onto integrated circuits. Electronics 38(8): 114–117

Mower J E (1992) Building a GIS for parallel computing environments  In Proceedings of the Fifth International Symposium on Spatial Data Handling 219–229. International Geographic Union, Columbia, SC

NIST (National Institute of Standards and Technology) (2015) *NIST Big Data Interoperability Framework: Volume 1, Definitions*. NIST Special Publication 1500-1. Gaithersburg, MD: NIST.
http://dx.doi.org/10.6028/NIST.SP.1500-1

NSF (National Science Foundation) (2018) CI2030: Future Advanced Cyberinfrastructure, A report of the NSF Advisory Committee for cyberinfrastructure
https://www.nsf.gov/cise/oac/ci2030/ACCI_CI2030Report_Approved_Pub.pdf

Ord JK, Getis A (1995) Local spatial autocorrelation statistics: distributional issues and an application. Geographical Analysis 27: 286–306

Rokos D, Armstrong MP (1996) Using Linda to compute spatial autocorrelation in parallel. Computers & Geosciences, 22 (5): 425-432

Sandu JS, Marble DF (1988) An investigation into the utility of the Cray X-MP supercomputer for handling spatial data. Proceedings of the Third International Symposium on Spatial Data Handling IGU Sydney, Australia pp. 253-266

Satyanarayanan M (2017) The emergence of edge computing. IEEE Computer 50(1): 30-39

Shekhar S, Ravada S, Kumar V, Chubb D, Turner G (1996) Parallelizing a GIS on a shared address space architecture. IEEE Computer 29 (12): 42-48

Shi W, Dustdar S (2016) The promise of edge computing. IEEE Computer 49(5): 78–81

Smarr L Catlett CE (1992) Metacomputing. Communications of the Association for Computing Machinery 35(6): 44–52

Snir M, Otto S, Huss-Lederman S, Walker D, Dongarra J (1996) MPI: The Complete Reference. Cambridge, MA: MIT Press.

Stewart C, Simms S, Plale B, Link M, Hancock D, Fox G (2010) What is Cyberinfrastructure? In: SIGUCCS '10 Proceedings of the 38th annual ACM SIGUCCS fall conference: navigation and discovery (Norfolk, VA, 24-27 Oct), pp. 37-44 https://dl.acm.org/citation.cfm?doid=1878335.1878347

Stone HS, Cocke, J (1991). Computer architecture in the 1990s. IEEE Computer 24(9): 30-38

Stonebraker M (1986) The case for shared nothing architecture. Database Engineering, Volume 9, Number 1

Sugumaran R, Armstrong MP (2017) Cloud computing. *The International Encyclopedia of Geography: People, the Earth, Environment, and Technology*. New York, NY: John Wiley
https://doi.org/10.1002/9781118786352.wbieg1017

Tang W (2017) GPU computing, edited by Michael F. Goodchild and Marc P. Armstrong, International Encyclopedia of Geography. https://doi.org/10.1002/9781118786352.wbieg0129

Tang W, Feng W (2017) Parallel map projection of vector-based big spatial data using general-purpose Graphics Processing Units. Computers, Environment and Urban Systems 61: 187-197

Wang S (2010) A CyberGIS framework for the synthesis of cyberinfrastructure, GIS, and spatial analysis. Annals of the Association of American Geographers 100(3): 535–557

Wang S (2013) CyberGIS: Blueprint for integrated and scalable geospatial software ecosystems. International Journal of Geographical Information Science 27(11): 2119–2121

Wang S (2019) Cyberinfrastructure. *The Geographic Information Science & Technology Body of Knowledge* (2nd Quarter 2019 Edition), John P. Wilson (Ed.) DOI: 10.22224/gistbok/2019.2.4

Wang S, Armstrong MP (2003) A quadtree approach to domain decomposition for spatial interpolation in grid computing environments. Parallel Computing 29(10): 1481–1504

Wang S, Armstrong MP (2009) A theoretical approach to the use of cyberinfrastructure in geographical analysis. International Journal of Geographical Information Science 23(2): 169–193 http://dx.doi.org/10.1080/13658810801918509

Wang S, Cowles MK, Armstrong MP (2008) Grid computing of spatial statistics: using the TeraGrid for Gi*(d) analysis. Concurrency and Computation: Practice and Experience 20 (14): 1697-1720 http://dx.doi.org/10.1002/cpe.1294

Yan J, Cowles MK, Wang S, Armstrong MP (2007) Parallelizing MCMC for Bayesian spatiotemporal geostatistical models. Statistics and Computing 17 (4): 323-335

Yang C, Huang Q, Li Z, Xu C, Liu K (2013) Spatial Cloud Computing: A Practical Approach. Boca Raton, FL: CRC Press